\documentclass[prd,nofootinbib,onecolumn]{revtex4}
\usepackage{amsmath}
\usepackage{amssymb}
\usepackage{graphics}
\usepackage{epsfig}

\newcommand\ba{\begin{eqnarray}}
\newcommand\ea{\end{eqnarray}}

\begin{document}

\title{Proton interactions with high multiplicity
}
\author{\firstname{A.~G.}~\surname{Afonin}}
\affiliation{%
IHEP, Protvino, Moscow region, Russia.
}%
\author{\firstname{A.~N.}~\surname{Aleev}}
\affiliation{%
JINR, Dubna, Moscow region, Russia.
}%
\author{\firstname{E.~N.}~\surname{Ardashev}}
\affiliation{%
IHEP, Protvino, Moscow region, Russia.
}%
\author{\firstname{V.~V.}~\surname{Avdeichikov}}
\affiliation{%
JINR, Dubna, Moscow region, Russia.
}%
\author{\firstname{V.~P.}~\surname{Balandin}}
\affiliation{%
JINR, Dubna, Moscow region, Russia.
}%
\author{\firstname{S.~G.}~\surname{Basiladze}}
\affiliation{%
SINP MSU, Moscow, Russia.
}%
\author{\firstname{M.~A.}~\surname{Batouritski}}
\affiliation{%
NC PHEP BSU, Minsk, Belarus.
}%
\author{\firstname{S.~F.}~\surname{Berezhnev}}
\affiliation{%
SINP MSU, Moscow, Russia.
}%
\author{\firstname{G.~A.}~\surname{Bogdanova}}
\affiliation{%
SINP MSU, Moscow, Russia.
}%
\author{\firstname{Yu.~T.}~\surname{Borzunov}}
\affiliation{%
JINR, Dubna, Moscow region, Russia.
}%
\author{\firstname{V.~A.}~\surname{Budilov}}
\affiliation{%
JINR, Dubna, Moscow region, Russia.
}%
\author{\firstname{Yu.~A.}~\surname{Chentsov}}
\affiliation{%
JINR, Dubna, Moscow region, Russia.
}%
\author{\firstname{V.~F.}~\surname{Golovkin}}
\affiliation{%
IHEP, Protvino, Moscow region, Russia.
}%
\author{\firstname{S.~N.}~\surname{Golovnya}}
\affiliation{%
IHEP, Protvino, Moscow region, Russia.
}%
\author{\firstname{S.~A.}~\surname{Gorokhov}}
\affiliation{%
IHEP, Protvino, Moscow region, Russia.
}%
\author{\firstname{N.~I.}~\surname{Grishin}}
\affiliation{%
SINP MSU, Moscow, Russia.
}%
\author{\firstname{Ya.~V.}~\surname{Grishkevich}}
\affiliation{%
SINP MSU, Moscow, Russia.
}%
\author{\firstname{G.~G.}~\surname{Ermakov}}
\affiliation{%
SINP MSU, Moscow, Russia.
}%
\author{\firstname{P.~F.}~\surname{Ermolov$ ^\dag $}}
\affiliation{%
SINP MSU, Moscow, Russia.
}%
\author{\firstname{N.~F.}~\surname{Furmanets}}
\affiliation{%
JINR, Dubna, Moscow region, Russia.
}%
\author{\firstname{D.~E.}~\surname{Karmanov}}
\affiliation{%
SINP MSU, Moscow, Russia.
}%
\author{\firstname{A.~V.}~\surname{Karpov}}
\affiliation{%
Department of Mathematics Komi SC UrD RAS, Syktyvkar, Russia.
}%
\author{\firstname{G.~D.}~\surname{Kekelidze}}
\affiliation{%
JINR, Dubna, Moscow region, Russia.
}%
\author{\firstname{V.~I.}~\surname{Kireev}}
\affiliation{%
JINR, Dubna, Moscow region, Russia.
}%
\author{\firstname{A.~A.}~\surname{Kiryakov}}
\affiliation{%
IHEP, Protvino, Moscow region, Russia.
}%
\author{\firstname{A.~G.}~\surname{Kholodenko}}
\affiliation{%
IHEP, Protvino, Moscow region, Russia.
}%
\author{\firstname{E.~S.}~\surname{Kokoulina}}
\email{kokoulin@sunse.jinr.ru}
\affiliation{%
JINR, Dubna, Moscow region, Russia.
}%
\author{\firstname{V.~V.}~\surname{Konstantinov}}
\affiliation{%
IHEP, Protvino, Moscow region, Russia.
}%
\author{\firstname{V.~N.}~\surname{Kramarenko}}
\affiliation{%
SINP MSU, Moscow, Russia.
}%
\author{\firstname{A.~V.}~\surname{Kubarovsky}}
\affiliation{%
SINP MSU, Moscow, Russia.
}%
\author{\firstname{A.~K.}~\surname{Kulikov}}
\affiliation{%
JINR, Dubna, Moscow region, Russia.
}%
\author{\firstname{E.~A.}~\surname{Kuraev}}
\affiliation{%
JINR, Dubna, Moscow region, Russia.
}%
\author{\firstname{L.~L.}~\surname{Kurchaninov}}
\affiliation{%
IHEP, Protvino, Moscow region, Russia.
}%
\author{\firstname{A.~Ya.}~\surname{Kutov}}
\affiliation{%
Department of Mathematics Komi SC UrD RAS, Syktyvkar, Russia.
}%
\author{\firstname{N.~A.}~\surname{Kuzmin}}
\affiliation{%
JINR, Dubna, Moscow region, Russia.
}%
\author{\firstname{G.~I.}~\surname{Lanschikov $ ^\dag $}}
\affiliation{%
JINR, Dubna, Moscow region, Russia.
}%
\author{\firstname{A.~K.}~\surname{Leflat}}
\affiliation{%
SINP MSU, Moscow, Russia.
}%
\author{\firstname{I.~S.}~\surname{Lobanov}}
\affiliation{%
IHEP, Protvino, Moscow region, Russia.
}%
\author{\firstname{E.~V.}~\surname{Lobanova}}
\affiliation{%
IHEP, Protvino, Moscow region, Russia.
}%
\author{\firstname{S.~I.}~\surname{Lutov}}
\affiliation{%
SINP MSU, Moscow, Russia.
}%
\author{\firstname{V.~N.}~\surname{Lysan}}
\affiliation{%
JINR, Dubna, Moscow region, Russia.
}%
\author{\firstname{M.~M.}~\surname{Merkin}}
\affiliation{%
SINP MSU, Moscow, Russia.
}%
\author{\firstname{G.~A.}~\surname{Mitrofanov}}
\affiliation{%
IHEP, Protvino, Moscow region, Russia.
}%
\author{\firstname{V.~V.}~\surname{Myalkovskiy}}
\affiliation{%
JINR, Dubna, Moscow region, Russia.
}%
\author{\firstname{V.~A.}~\surname{Nikitin}}
\affiliation{%
JINR, Dubna, Moscow region, Russia.
}%
\author{\firstname{V.~D.}~\surname{Peshehonov}}
\affiliation{%
JINR, Dubna, Moscow region, Russia.
}%
\author{\firstname{V.~S.}~\surname{Petrov}}
\affiliation{%
IHEP, Protvino, Moscow region, Russia.
}%
\author{\firstname{Y.~P.}~\surname{Petukhov}}
\affiliation{%
JINR, Dubna, Moscow region, Russia.
}%
\author{\firstname{A.~V.}~\surname{Pleskach}}
\affiliation{%
IHEP, Protvino, Moscow region, Russia.
}%
\author{\firstname{M.~K.}~\surname{Polkovnikov}}
\affiliation{%
IHEP, Protvino, Moscow region, Russia.
}%
\author{\firstname{V.~V.}~\surname{Popov}}
\affiliation{%
SINP MSU, Moscow, Russia.
}%
\author{\firstname{V.~N.}~\surname{Riadovikov}}
\affiliation{%
IHEP, Protvino, Moscow region, Russia.
}%
\author{\firstname{V.~N.}~\surname{Ronzhin}}
\affiliation{%
IHEP, Protvino, Moscow region, Russia.
}%
\author{\firstname{I.~A.}~\surname{Rufanov}}
\affiliation{%
JINR, Dubna, Moscow region, Russia.
}%
\author{\firstname{V.~A.}~\surname{Senko}}
\affiliation{%
IHEP, Protvino, Moscow region, Russia.
}%
\author{\firstname{N.~A.}~\surname{Shalanda}}
\affiliation{%
IHEP, Protvino, Moscow region, Russia.
}%
\author{\firstname{M.~M.}~\surname{Soldatov}}
\affiliation{%
IHEP, Protvino, Moscow region, Russia.
}%
\author{\firstname{V.~I.}~\surname{Spiryakin}}
\affiliation{%
JINR, Dubna, Moscow region, Russia.
}%
\author{\firstname{A.~V.}~\surname{Terletskiy}}
\affiliation{%
JINR, Dubna, Moscow region, Russia.
}%
\author{\firstname{L.~A.}~\surname{Tikhonova}}
\affiliation{%
SINP MSU, Moscow, Russia.
}%
\author{\firstname{Yu.~P.}~\surname{Tsyupa}}
\affiliation{%
IHEP, Protvino, Moscow region, Russia.
}%
\author{\firstname{A.~M.}~\surname{Vishnevskaya}}
\affiliation{%
SINP MSU, Moscow, Russia.
}%
\author{\firstname{V.~Yu.}~\surname{Volkov}}
\affiliation{%
SINP MSU, Moscow, Russia.
}%
\author{\firstname{A.~P.}~\surname{Vorobiev}}
\affiliation{%
IHEP, Protvino, Moscow region, Russia.
}%
\author{\firstname{A.~G.}~\surname{Voronin}}
\affiliation{%
SINP MSU, Moscow, Russia.
}%
\author{\firstname{V.~I.}~\surname{Yakimchuk}}
\affiliation{%
IHEP, Protvino, Moscow region, Russia.
}%
\author{\firstname{A.~I.}~\surname{Yukaev}}
\affiliation{%
JINR, Dubna, Moscow region, Russia.
}%
\author{\firstname{V.~N.}~\surname{Zapolskii}}
\affiliation{%
IHEP, Protvino, Moscow region, Russia.
}%
\author{\firstname{N.~K.}~\surname{Zhidkov}}
\affiliation{%
JINR, Dubna, Moscow region, Russia.
}%
\author{\firstname{S.~A.}~\surname{Zotkin}}
\affiliation{%
SINP MSU, Moscow, Russia.
}%
\author{\firstname{E.~G.}~\surname{Zverev}}
\affiliation{%
SINP MSU, Moscow, Russia.
}%

\date{\today}

\begin{abstract}
Project Thermalization (Experiment SERP-E-190 at IHEP) is aimed to
study the proton - proton interactions at 50 GeV with large number
of secondary particles. In this report the experimentally measured
topological cross sections are  presented taking into account the
detector response and procession efficiency. These data are in good
agreement with gluon dominance model. The comparison with other
models is also made and shows no essential discrepancies.

{$^\dag $ deceased}
\end{abstract}

\maketitle

\section{Introduction}
The study of high multiplicity processes is closely connected with
understanding of the nature of strong interactions. The project
Thermalization (experiment SERP-E-90 at IHEP) \cite {Avdeichikov} is
aimed to study of events with multiplicity significantly exceeding
the average one. We carry out the detection of these unique events
at U-70 accelerator (IHEP, Protvino) in 50~GeV proton beam. The main
aim of the project is to study the collective behavior of secondary
particles in the extreme multiplicity region.

In seventies Mirabelle bubble chamber Collaboration had measured the
multiplicity up to 16 charged particles in pp interactions at 50 GeV
\cite{6}. SVD Collaboration continues the search for the events with
multiplicity more than 20 both charged and neutral particles. To
reach the goal, we have renewed SVD-2 setup (Fig. 1) at  U-70
accelerator of IHEP (Protvino). Now it is equipped with a liquid
hydrogen target \cite{Golov} micro-strip silicon detector, a
magnetic spectrometer with proportional chambers \cite{MS}, a drift
tube tracker \cite{Basil}, Cherenkov counter, electromagnetic
calorimeter for registration of gammas \cite{Kir}.

To suppress registration of the events with low charged
multiplicity, we have implemented the special scintillation
hodoscope for triggering \cite{Trig, NPCS}. Using this setup we have
extended the charged particle multiplicity measurements from 16
(Mirabelle data) to 24 particles at present. The achieved value of
the smallest topological cross section is less by three orders of
magnitude in comparison with the Mirabelle results. Measured charged
particle multiplicity distribution has been corrected for apparatus
acceptance and detection efficiency and compared with some models
predictions.

\begin{figure}
\includegraphics[width=6.2 in, height=3.1 in, angle=0]{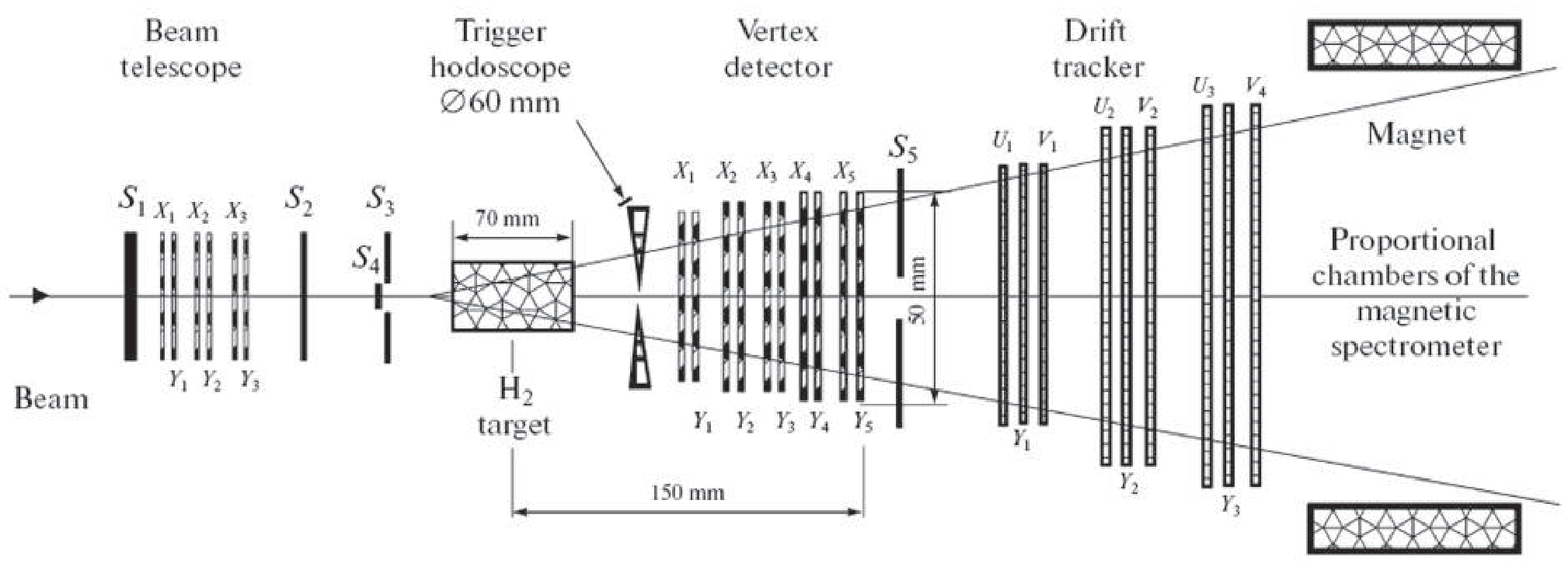}
\caption{}{Schematic diagram of the SVD-2 setup.}
\end{figure}

The collective behavior of secondary particles is expected to onset
in the extreme multiplicity region. In particular, it may evidence
for the Bose-Einstein condensation which has been predicted in this
area. The calculation by the MC PYTHIA code has shown that the
standard generator predicts the values of the topological cross
section at 70 GeV (the energy of U-70) which is in a reasonably good
agreement with the experimental data at small multiplicity
($n_{ch}<10$)  but it underestimates the value $\sigma(n_{ch})$ by
two orders of magnitude at $n_{ch}=18$ \cite{Avdeichikov}.

\begin{figure}
\includegraphics[width=3.9 in, height=3.5 in, angle=0]{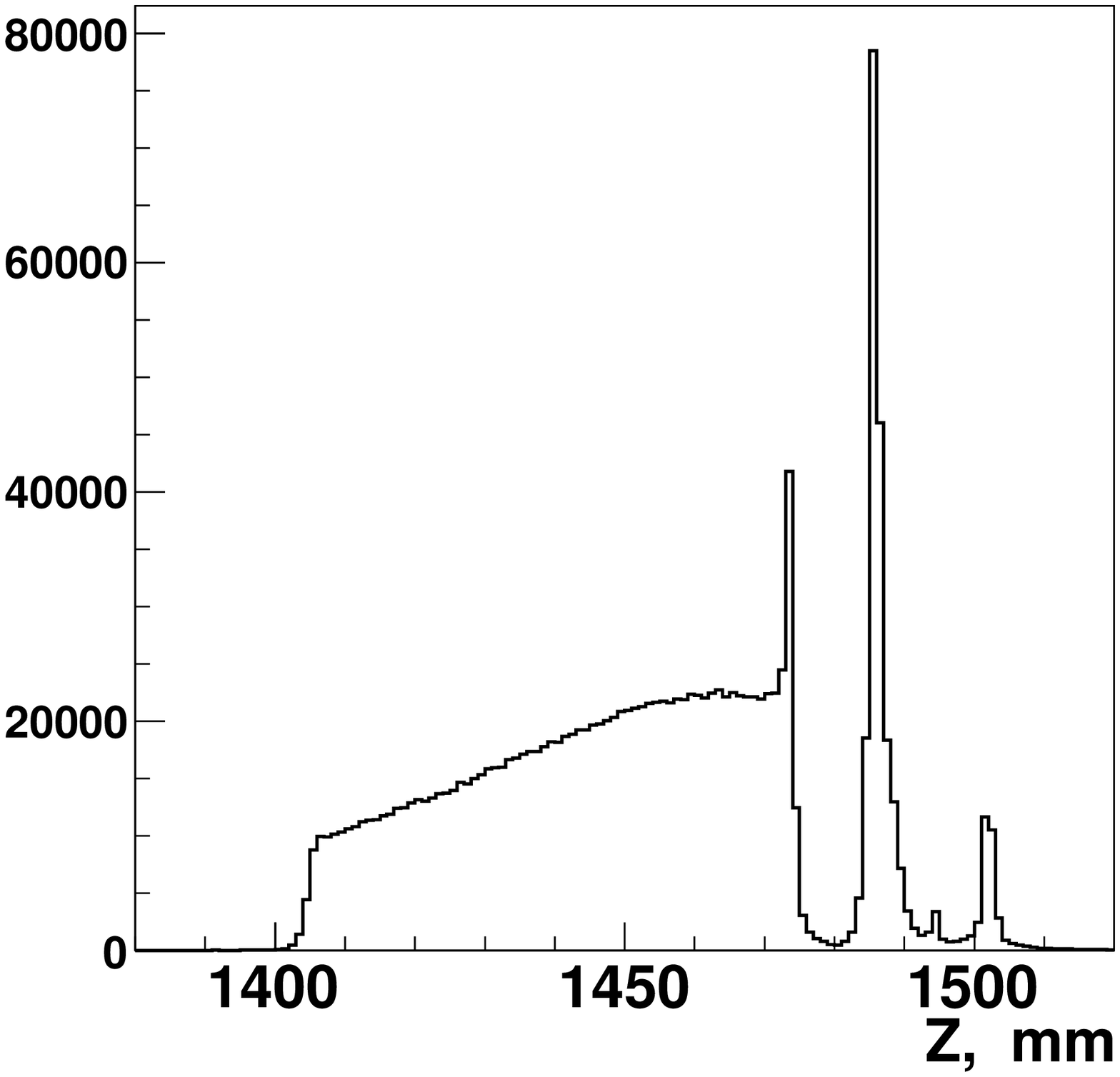}
\caption{}{The distribution on \emph{Z} -coordinate of the
interaction vertex in the hydrogen target. }
\end{figure}

\section{Event selection, Track Fitting and Correction Procedure}
The main element of SVD-2 setup is a micro-strip silicon vertex
detector with 10 planes. It allows the reconstruction of the
interaction vertex and tracks. We have obtained the multiplicity
distribution for this report using vertex detector data only. The
5.13 millions of events were taken during 2008 year run of SVD
setup. From this statistics 3.85 millions of events have been taken
at trigger-level 8 (lower limit of the multiplicity set at trigger
system). Out of them 2.1 millions of events have been detected in
the fiducial volume of the hydrogen target. For final analysis 1.0
millions of events were remained. They were selected according to
the criterions:

a) the number of beam tracks simultaneously hitting the target is
not exceed 2;

b) the uncertainty of the vertex reconstruction on two projections
is smaller than 5 mm.

\begin{figure}
\includegraphics[width=3.1 in, height=3.0 in, angle=0]{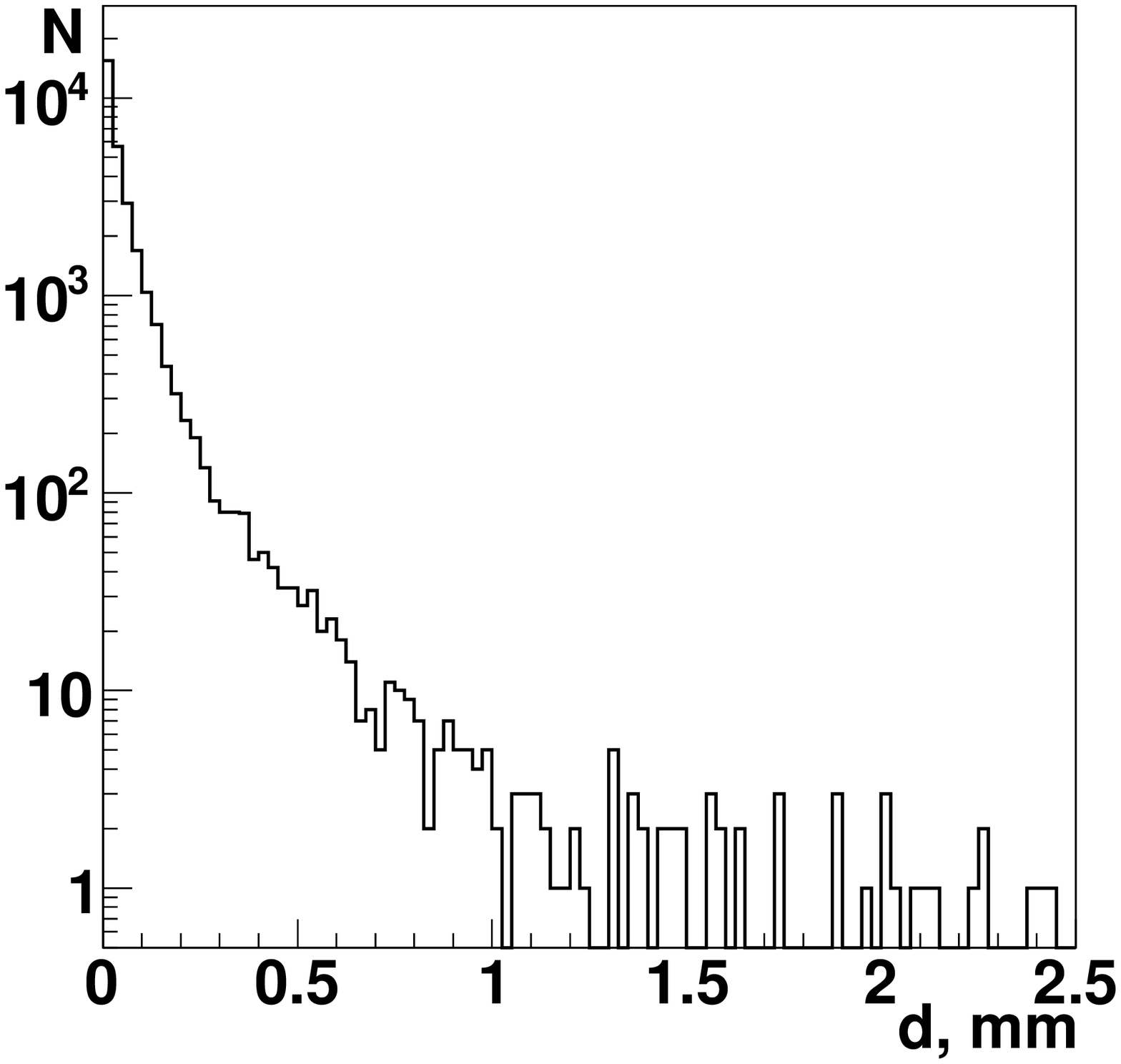}
\caption{}{The distribution of simulated tracks on impact. }
\end{figure}

\begin{figure}
\includegraphics[width=3.1 in, height=3.0 in, angle=0]{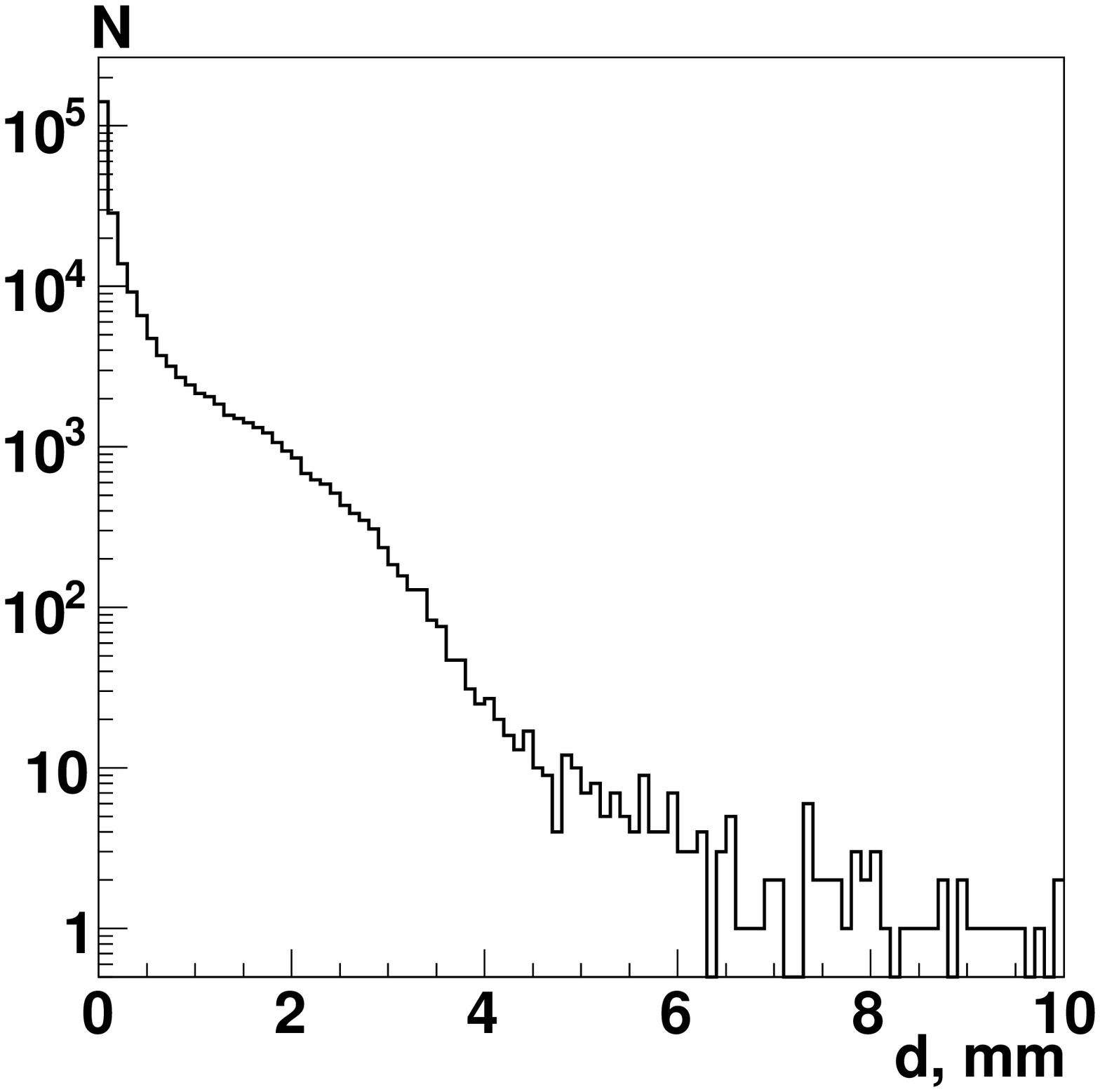}
\caption{}{The distribution of the experimental tracks on impact.}
\end{figure}
The distribution on reconstructed \emph{X} -coordinate for the
interaction vertex in the hydrogen target is obtained. \emph{X}
-coordinate axis, as \emph{Y} -coordinate axis, are directed to
perpendicular of beam direction. The distribution on \emph{Y}
-coordinate is differed from \emph{X} -coordinate distribution
insignificantly. The distribution on \emph{Z} -coordinate of the
vertex interaction in the hydrogen target is presented in Fig.~2.
\emph{Z} -coordinate axis is directed to the proton beam. The
interval 1405 - 1470 (mm) corresponds to interactions in the
hydrogen target. Peaks on the right side appear from interactions in
film window, shell of the target and scintillation hodoscope.

The reconstruction algorithm was taken as following. The space
points of the hits are reconstructed accounting on one- and
multi-strip clusters. The center of gravity method was applied for
coordinate evaluation:  $\overline X = \sum\limits_i A_i \times
X_i/\sum\limits_i A_i,$ where $X_i$ is a strip coordinate and $A_i$
--- a strip signal amplitude. A track has been approximated by a straight
line using separately the \emph{X} and \emph{Y} coordinates with 3
or 4 hits. For track reconstruction the Kalman filter has been
applied. It effectively rejects the random noise on strip planes.
The length of the target is 70 mm. The initial approximate location
of primary vertex is assigned in a half of the target closest to the
vertex detector. This is done to minimize apparatus acceptance
correction. At least the presence of the one hit is required on the
first two micro-strip planes after a target. Among a few tracks
candidates the candidate with the best least chi-square fit is
selected. If one candidate with four hits has worse chi-square value
than candidates with three hits then the candidates with four hits
is taken. In our method two tracks can pass through common (one or
two) hits.

Multiple scattering in vertex detector causes the track deviation
from straight line. In our case this deviation in position of last
plane does not exceed the coordinate precision measurements. The
vertex of interaction is determined by the least-squares method. The
tracks with high deviation from vertex are not included in the
event. The simulation is shows that the vertex interaction
reconstruction precision amounts 30 mcm on \emph{X} and \emph{Y}
axes, 400 mcm on \emph{Z} axis without taking into alignment
uncertainty. The experimental values of impact for the vertex
position measurement are equal to 0.28 mm on \emph{X} and 0.36 mm on
\emph{Y} coordinates. The absent of correct alignment procedure
makes worse the coordinate precision determination greater than 3
times.

\begin{figure}
\includegraphics[width=3.5 in, height=3.2 in, angle=0]{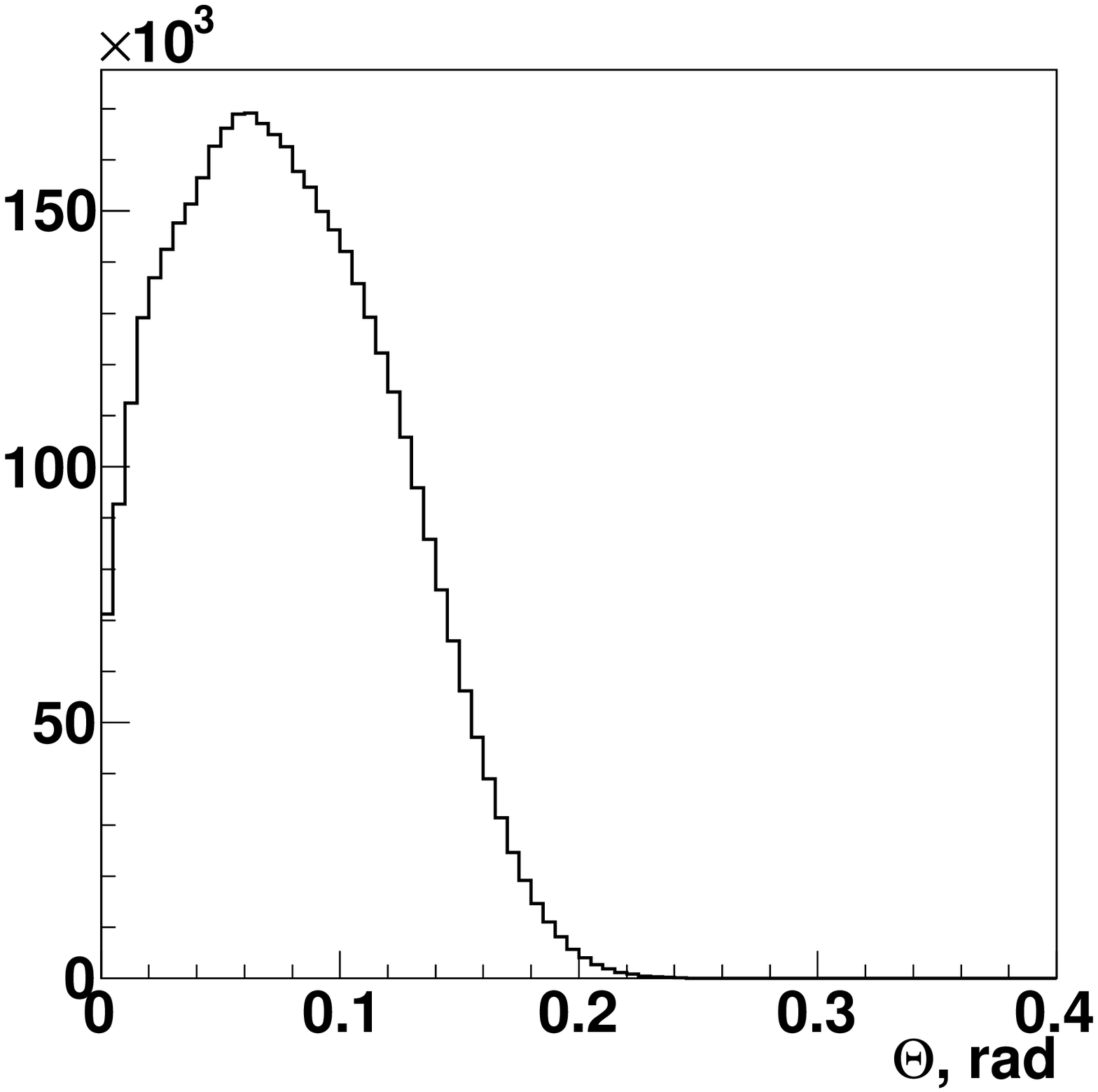}
\caption{}{The angular distribution of reconstructed tracks on the
polar angle $\Theta $ on experimental data.}
\end{figure}

The simulation shows that the number of tracks deviated from the
vertex more then 1 mm amounts 0.1\% (Fig.~3). In our experiment the
number of such tracks is equal to 9\% (Fig.~4). These tracks come
from secondary vertex or are fakes. Their sources are noise,
misalignment and secondary interactions. Tracks in the space are
reconstructed by means of two oblique planes $\it U$ and $\it V$
located at the end of vertex detector. The angular distribution of
reconstructed tracks from the initial vertex on the polar angle
$\Theta $ for all multiplicities is presented in Fig.~5.

The correction procedure of charged multiplicity distributions is
carried out taking into account an influence of the multiplicity
trigger conditions and inefficiency of track reconstruction
algorithm and acceptance of vertex detector. To make these
corrections we used tables of spread coefficients on multiplicity,
$a_{ij}=N_i/N_j,$ where $a_{ij}$ is the probability to reconstruct
successfully $i$ charged tracks for event with $j$ charged tracks,
$N_j$ --- the number of simulated events with $j$ charged tracks,
from which $N_i$ events were reconstructed as events with $i$
charged tracks. The index $i$ is changed from 1 up to 24, the index
$j$ takes only even values from 2 up to 24. The table of
coefficients is calculated using Monte Carlo simulation (GEANT3).
This procedure is used to calculate the acceptance of the apparatus
along with the reconstruction and triggering efficiencies.

We get the overdetermined system of linear equations, in general
case 24 equations with 12 unknown quantities $x_j$: $
\sum\limits_{j=2}^{24} a_{ij} x_j = b_i,$ where $b_i$ is the
experimental number of events with multiplicity $i$. This system can
be solved by the ordinary Gauss-Seidel method or by the
least-squares method \cite{Zaidel}. It is difficult to account for
the trigger inefficiency below it threshold (8 minimum ionizing
particles, MIP) so we publish here corrected topological cross
section for $n_{ch}\geq 10$ where trigger efficiency is close to 1
and its influence is insignificant. For low multiplicity ($n_{ch}
\leq 10$) we use MIRABELLE data for absolute normalization of our
cross section. The event simulation for high multiplicity is carried
out in accordance with Boltzmann and Bose models  and taking into
account the acceptance and the efficiency of the vertex detector
{\cite{Avdeichikov}. The differences in $a_{ij}$ coefficients
obtained for these two models were used to estimate the systematical
errors of the correction coefficients. In the Table~1 we give
topological cross section obtained by Mirabelle Collaboration
\cite{6}. The corrected topological cross sections for pp
interactions at 50 GeV with statistical errors (the negligible
systematical errors are not included) are presented in the Table~2.\\\

\begin{figure}
\includegraphics[width=4.9 in, height=3.8 in, angle=0]{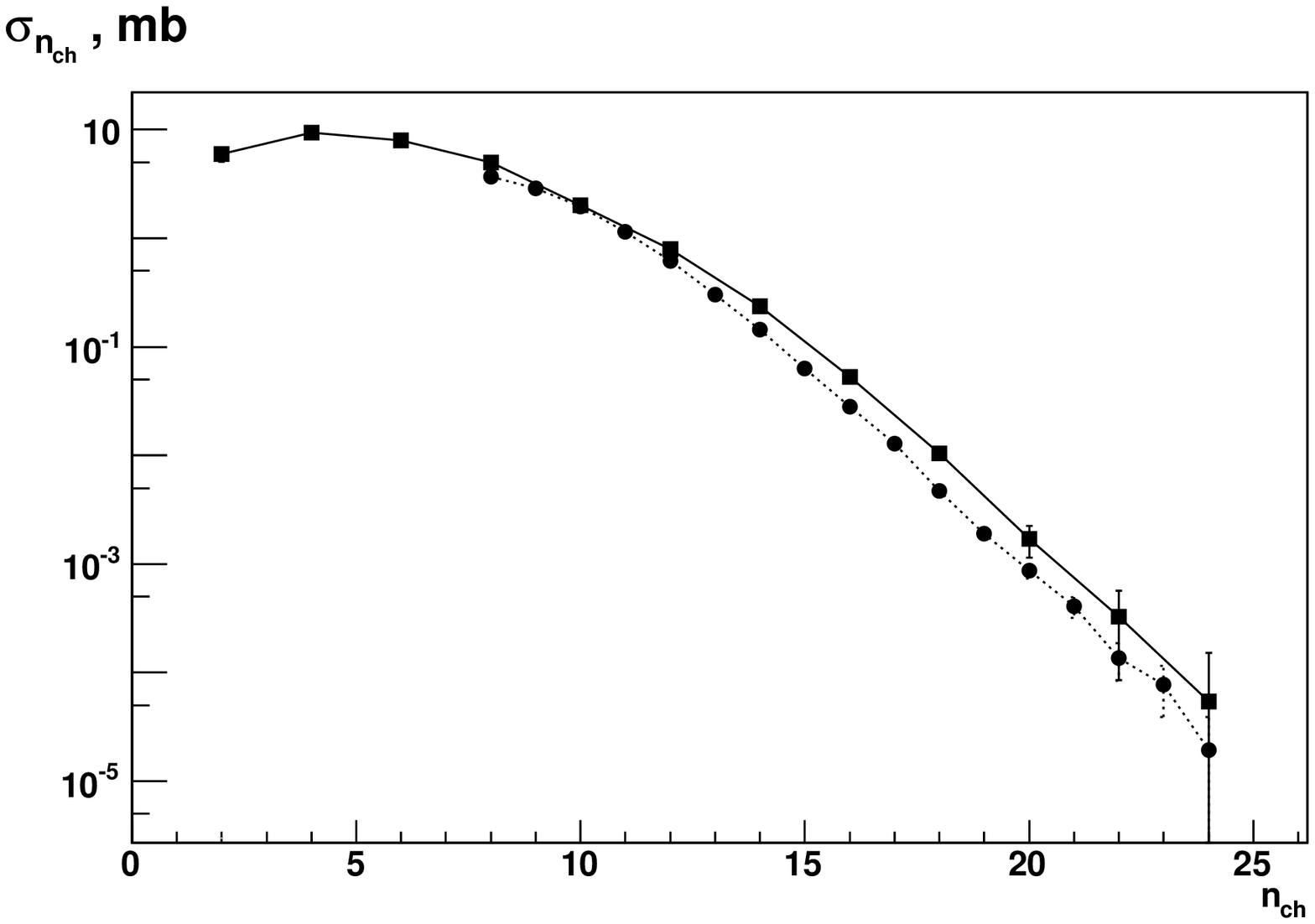}
\caption{}{Experimental topological cross sections for pp
interactions with Mirabelle data~\cite{6} addition before (filled
circles)) and after (filled boxes) the inclusion of corrections.
Only statistical errors are plotted. }
\end{figure}

Table. 1. The topological cross sections at 50 GeV in pp
interactions obtained by Mirabelle Collaboration \cite{6}.

\begin{center}
\begin{tabular}{||c||c|c|c|c|c|c|c|c||}
\hline \hline
$n_{ch}$& 2 & 4 & 6 & 8  & 10  & 12 & 14  & 16 \\
\hline
$\sigma (n_{ch})$  & 5.97  & 9.40     & 7.99 & 5.02 & 2.03  & 0.48 & 0.20 & 0.01 \\
\hline
$\Delta \sigma (n_{ch})$  & 0.88 & 0.47     & 0.43 & 0.33 & 0.20  & 0.10 & 0.06 & 0.02 \\
\hline\hline
\end{tabular}
\end{center}

Table. 2. The topological cross sections obtained by SVD
Collaboration in pp interactions at 50 GeV.
\begin{center}
\begin{tabular}{||c||c|c|c|c|c|c|c|c||}
\hline \hline
$n_{ch}$& 10 & 12 & 14& 16  & 18  & 20 & 22  & 24 \\
\hline
$\sigma (n_{ch})$  & 1.685  & 0.789     & 0.234 & 0.0526 & 0.0104  &0.0017 & 0.00033 & 0.000054\\
\hline
$\Delta \sigma (n_{ch})$  & 0.017 & 0.012     & 0.006  & 0.0031 & 0.0014  &0.0006 & 0.00024 & 0.000098\\
\hline\hline

\end{tabular}
\end{center}

\begin{figure}
\includegraphics[width=4.9 in, height=3.8 in, angle=0]{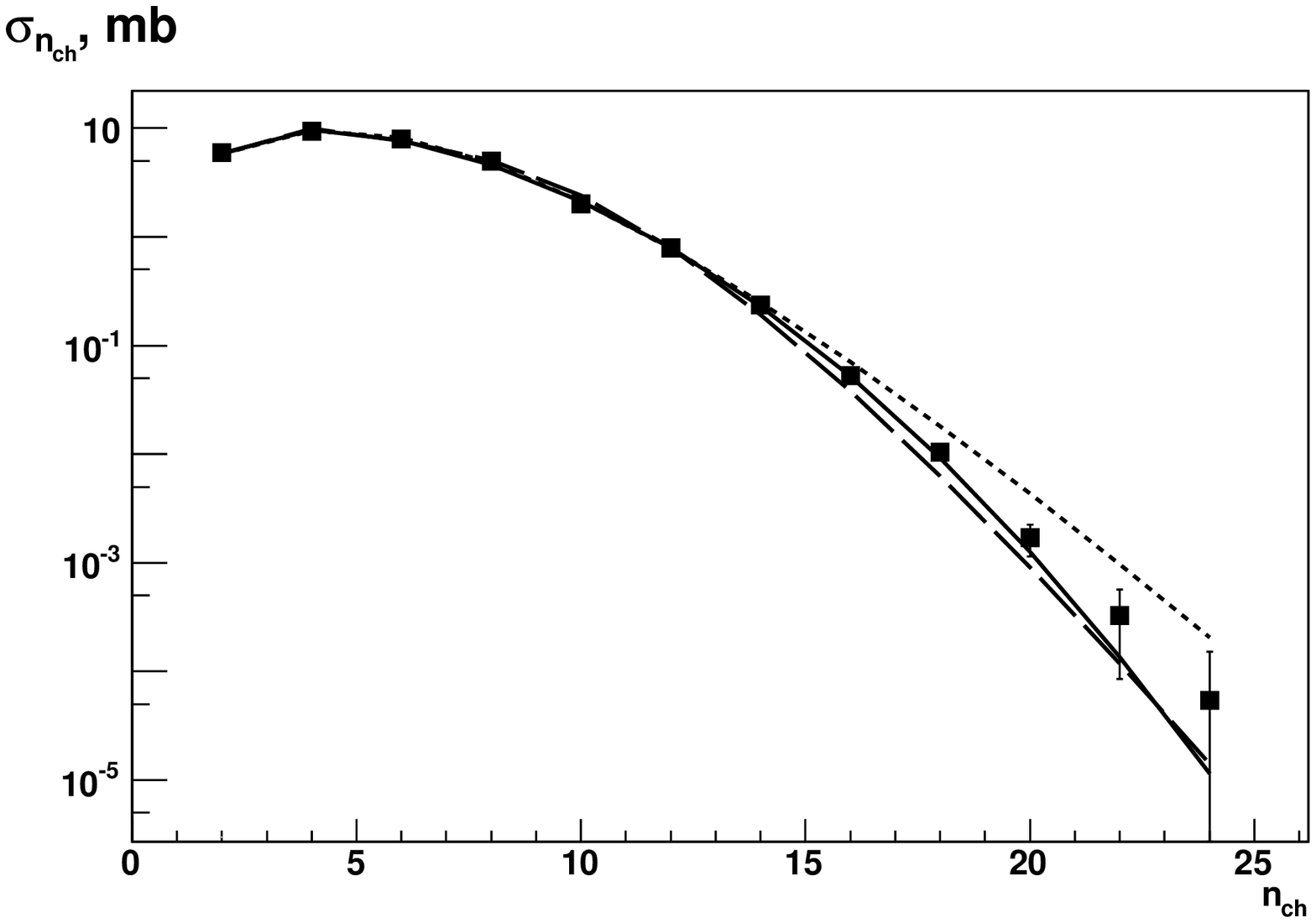}
\caption{}{Comparisons data with GDM (solid curve), IHEP model
(dotted curve) and NBD distribution (dashed curve).}
\end{figure}

We have renovated previous Mirabelle data \cite{6} for $n_{ch}$ from
10 up 16, and added 4 new points from 18 up to 24. The cross section
at the last point, $n_{ch}=24$, is three order of magnitude lower
than previously known cross section at $n_{ch} = 16$ \cite{6}. We
normalized our data to Mirabelle one \cite{6} in the region
$n_{ch}$=10--16. This allows us to get inelastic cross section,
$\sigma (n_{ch}) = 31.50 \pm 1.14$ (mb) at 50 GeV and the mean
charged multiplicity, $\overline n(s)=5.45 \pm 0.24$. We also
calculated the variance, $D_2=7.21 \pm 2.80$ and second correlative
moment, $f_2$ = 1.75. Available uncorrected experimental data at
proton energy 50 GeV are shown in Fig.~6 (filled circles). Corrected
data on detector response are shown in Fig.~6 too (filled boxes).

\section{Comparisons with models}
The comparison of topological cross sections with three models is
carried out (Fig.~7). Statistical errors are plotted (the negligible
systematical errors are not shown).

Now there are only few phenomenological models giving predictions on
the multiplicity distributions at the extreme domain
\cite{chikilev,GDM,Tyur,Giov}.  One of them is why the gluon
dominance model (GDM) \cite{GDM} has been developed. It is based on
the main essences of QCD and supplemented with the phenomenological
mechanism of hadronization. This approach shows the activity of
gluons and the passive role of the quarks in the multiparticle
production mechanism. GDM confirms convincingly the recombination
mechanism of hadronization in hadron and nuclear interactions and
fragmentation in lepton processes. In Fig.~7 the description by GDM
is presented by solid curve. The essence of the GDM is the
convolution of the multiplicity distributions on two stages: the
parton (gluon) cascade and hadronization described by
phenomenological scheme. The active gluons play dominant role in
multiparticle production of hadrons \cite{GDM}.

An analytical expression for multiplicity distribution in the
KNO-form was obtained by a theoretical group from IHEP \cite{Tyur}
at seventies. This model has combined the inelastic and elastic
processes at high energies using spectral densities of inelastic
channel contributions into unitarity condition obtained at
stochastic description of collisions at high energy. It permits to
get multiplicity distributions which gives good agreement to
experimental data. Comparison of this function with our data is
presented in Fig.~7 by the dotted curve. Evidently the agreement
with the data is good.

The negative binomial distribution (NBD) \cite{Giov} is the commonly
utilized formula for multiplicity distributions. This distribution
is obtained in clan structure of interactions and is manifested in
two stage dynamical process of multiparticle production. At first
stage the germs (parents) are formed and then cascades are
produced.The comparison of this model with data is shown in Fig.~7
by the dashed curve. It does not describe well the data in the high
multiplicity region (it exceeds our data).

These investigations have been partially supported by Russian
Foundation of Basic Research nos. $08-02-90028-\rm {Bel\_a}$ ,
$09-02-92424-\rm {KE}\_a$,09-02-00445a, 06-02-16954, president of
Russia grant for leading scientific school support 1456.2008.2. We
appreciate to IHEP leadership for support in the carrying out of
investigations,thank Operations Group and department of beams
provided effective work of U-70 and 22 channel.

%
%
%
%
%
%
%
%
%
%

\end{document}